\documentclass[prl,aps,amsmath,twocolumn,floatfix,showpacs,superscriptaddress]{revtex4}

\usepackage{bm}
\usepackage{graphicx}
\usepackage{multirow}
\usepackage{bm}

\begin{document}
\title{Process tomography of ion trap quantum gates}

\author{M.~Riebe}
\author{K.~Kim}
\author{P.~Schindler}
\author{T.~Monz}
\affiliation{Institut f{\"u}r Experimentalphysik, Universit{\"a}t Innsbruck,
Technikerstrasse 25, A-6020 Innsbruck, Austria}

\author{P.~O.~Schmidt}
\affiliation{Institut f{\"u}r Experimentalphysik, Universit{\"a}t Innsbruck,
Technikerstrasse 25, A-6020 Innsbruck, Austria}

\author{T.~K.~K\"orber}
\affiliation{Institut f{\"u}r Experimentalphysik, Universit{\"a}t
Innsbruck, Technikerstrasse 25, A-6020 Innsbruck, Austria}
\affiliation{Institut f\"ur Quantenoptik und Quanteninformation
der \"Osterreichischen Akademie der Wissenschaften, Technikerstr.
21a, A-6020 Innsbruck, Austria}

\author{W.~H{\"a}nsel}
\affiliation{Institut f{\"u}r Experimentalphysik, Universit{\"a}t Innsbruck,
Technikerstrasse 25, A-6020 Innsbruck, Austria}

\author{H.~H{\"a}ffner}
\affiliation{Institut f{\"u}r Experimentalphysik, Universit{\"a}t
Innsbruck, Technikerstrasse 25, A-6020 Innsbruck, Austria}
\affiliation{Institut f\"ur Quantenoptik und Quanteninformation
der \"Osterreichischen Akademie der Wissenschaften, Technikerstr.
21a, A-6020 Innsbruck, Austria}

\author{C.~F.~Roos}
\affiliation{Institut f{\"u}r Experimentalphysik, Universit{\"a}t Innsbruck,
Technikerstrasse 25, A-6020 Innsbruck, Austria} \affiliation{Institut f\"ur
Quantenoptik und Quanteninformation der \"Osterreichischen Akademie der
Wissenschaften, Technikerstr. 21a, A-6020 Innsbruck, Austria}

\author{R.~Blatt}
\affiliation{Institut f{\"u}r Experimentalphysik, Universit{\"a}t
Innsbruck, Technikerstrasse 25, A-6020 Innsbruck, Austria}
\affiliation{Institut f\"ur Quantenoptik und Quanteninformation
der \"Osterreichischen Akademie der Wissenschaften, Technikerstr.
21a, A-6020 Innsbruck, Austria}

\date{\today}

\begin{abstract}
A crucial building block for quantum information processing with trapped ions
is a controlled-NOT quantum gate. In this paper, two different sequences of
laser pulses implementing such a gate operation are analyzed using quantum
process tomography. Fidelities of up to $92.6(6)\:\%$ are achieved for single
gate operations and up to $83.4(8)\:\%$ for two concatenated gate operations.
By process tomography we assess the performance of the gates for different
experimental realizations and demonstrate the advantage of amplitude--shaped
laser pulses over simple square pulses. We also investigate whether the
performance of concatenated gates can be inferred from the analysis of the
single gates.
\end{abstract}


\maketitle Processing information with well-controlled quantum
systems has the fascinating perspective of being much more
powerful than classical computers for certain applications. A
promising candidate for the experimental realization of quantum
computing are strings of ions stored in linear Paul traps
\cite{CiracZollerPRL1995} as recently demonstrated by various key
experiments, including the preparation of multi-particle entangled
states \cite{haeffner2005,leibfried2005}, quantum teleportation
\cite{riebe2004,barrett2004} and quantum error correction
\cite{chiaverini2004}. Quantum information processing depends on
the ability to implement single qubit rotations and most
importantly an entangling two-qubit quantum gate
\cite{schmidtkalernature2003,leibfried2003,haljan2005,home2006}.
Proper characterization and understanding of the action of gate
operations and their imperfections is of vital importance in order
to successfully apply them in complex computations.

Generally, the implementations of quantum gates are imperfect due
to decoherence and various systematic error sources present in
experimental setups. A proper description of such an operation
which accounts for the possibly non-unitary evolution of the
qubits is provided by quantum process tomography
\cite{Chuang1997,poyatos1997}. Process tomography has already been
applied for characterizing quantum gates in NMR and linear-optics
quantum computing \cite{childs2001,obrien2004,kiesel2005}. Here,
we show that process tomography is a valuable tool for comparing
different ion trap quantum gate implementations and optimizing the
experimental parameters. This way, we were able to improve our
controlled-NOT (CNOT) gate fidelity from 71 \%
\cite{schmidtkalernature2003} to almost 93 \%. Moreover, the
action of two successively applied gate operations is investigated
and compared to the predictions from the single gate tomography
result.

We realize entangling gates between $^{40}$Ca$^+$ ions held in a linear trap
\cite{schmidtkalerapplphysb2003}. Quantum information is stored in
superpositions of the $|S\rangle\equiv S_{1/2}(m=-1/2)$ ground state and the
metastable $|D\rangle\equiv D_{5/2}(m=-1/2)$ state and is manipulated by laser
pulses at a wavelength of 729 nm exciting the electric quadrupole transition
between those states. A focus size smaller than the inter-ion distance and
precise control of the focus position allows us to address single qubits.
Detection of the qubit's quantum state is achieved by scattering light on the
$S_{1/2}\leftrightarrow P_{1/2}$ dipole transition and detecting the presence
or absence of resonance fluorescence of the individual ions with a CCD-camera.
The oscillations of the ions in the harmonic trap potential are described by
normal modes and give rise to sidebands in the spectrum of the
$S_{1/2}\leftrightarrow D_{5/2}$ transition. For coherent state manipulation,
only the quantum states $|n\rangle$ of the axial center-of-mass mode at a
frequency $\omega_z=2\pi\cdot 1.36$ MHz are relevant. Here, $n$ denotes the
number of vibrational quanta. Quantum information processing is implemented by
(a) laser pulses on the carrier transitions $|S,n\rangle \leftrightarrow
|D,n\rangle$ realizing single qubit operations on the ion qubits and (b) laser
pulses on the first blue sideband inducing transitions between the states
$|S,n\rangle$ and $|D,n+1\rangle$ which connect the internal state of the ions
and the state of the vibrational mode. The latter operation allows us to
implement an entangling interaction between ion qubits
\cite{CiracZollerPRL1995}. A more detailed account of our experimental setup
can be found in Ref.\ \cite{schmidtkalerapplphysb2003}.

Cirac-Zoller controlled-NOT gate operations between two ion qubits are
implemented by the pulse sequences shown in Tab.\ \ref{tab:czpulses}. In both
sequences,  the quantum state of the control qubit 2 is first mapped to the
vibrational mode (SWAP I), which is cooled to its ground state prior to the
operation. Then a CNOT operation is performed between the vibrational mode and
the target ion qubit 1. Finally the state of the vibrational mode is mapped
back onto the control qubit (SWAP II), restoring its quantum state and
returning the vibrational mode to its ground state. Both pulse sequences differ
only in the way the phase gate sequence is implemented. The ideal unitary
evolution realized by the first pulse sequence (A) is
\cite{schmidtkalernature2003}
\begin{align}
&U_{\text{CNOT}}^{\text{(A)}}=\left(\begin{array}{cccc}
  0 &i  &0  &0  \\
  -i & 0 & 0 & 0 \\
  0 & 0 & -1 & 0 \\
  0 & 0 & 0 & -1 \\
\end{array}\right)\notag \\
\label{eq:cnotunitaryA} &=-\frac{1}{2}\left(\hat{I}\otimes
\hat{I}-\hat{Z}\otimes \hat{I}+\hat{I}\otimes \hat{Y}+\hat{Z}\otimes
\hat{Y}\right),
\end{align}
where the matrix is written in the product basis
$\{|DD\rangle,|DS\rangle,|SD\rangle,|SS\rangle\}$ and expressed in terms of the
Pauli operators $\hat{X}$, $\hat{Y}$, $\hat{Z}$ and the identity $\hat{I}$. In
this sequence the state of target ion 1 is flipped whenever control ion 2 is in
state $|D\rangle$ (the order of the ions is $|\text{ion 2},\text{ion
1}\rangle$). The unitary evolution of the second pulse sequence (B) is
\cite{childspra2000}
\begin{align}
&U_{\text{CNOT}}^{\text{(B)}}= \hat{U}_Z\cdot\left(
\begin{array}{cccc}
  1 &0  &0  &0  \\
  0 & 1 & 0 & 0 \\
  0 & 0 & 0 & i \\
  0 & 0 & -i & 0
\end{array}\right)\notag \\
&= \frac{1}{2}\:\hat{U}_Z\cdot\left(\hat{I}\otimes \hat{I}+\hat{Z}\otimes
\hat{I}-\hat{I}\otimes \hat{Y}+\hat{Z}\otimes
\hat{Y}\right).\label{eq:cnotunitaryB}
\end{align}
Here the state of the target ion 1 is flipped whenever the control
ion 2 is in $|S\rangle$. This pulse sequence shows the desired
unitary evolution if an additional rotation
$\hat{U}_Z=\exp\left[-i\cdot(1-1/\sqrt{8})\pi\cdot\hat{Z}\right]\otimes
\exp\left[-i\cdot\pi/\sqrt{8}\cdot\hat{Z}\right]$ is applied. If
this CNOT pulse sequence is embedded in a larger algorithm, the
additional rotation can be taken into account by shifting the
phase of every subsequent pulse by
$\Delta\phi=-1/\sqrt{2}\cdot\pi$ on the control ion and by
$\Delta\phi=+1/\sqrt{2}\cdot\pi$ on the target ion. As can be seen
from Tab.\ \ref{tab:czpulses}, pulse sequence (B) is shorter in
terms of total length of the sideband pulses than sequence (A).
\begin{table}[hbt]
\begin{tabular}{c|c|c}
Description& Seq. (A) & Seq. (B)\\
\hline
SWAP I& $R^+_2(\pi,0)$&$R^+_2(\pi,0)$ \\
\hline
Ramsey I& $R_1(\pi/2,0)$& $R_1(\pi/2,0)$\\
\hline \multirow{4}*{Phase gate}
 &$R_1^+(\pi,\pi/2) $&$R_1^+(\pi/2,\pi)$\\
&$R_1^+(\pi/\sqrt{2},0) $& $R_1^+(\sqrt{2}\pi,\pi/2)$\\
&$R_1^+(\pi,\pi/2) $&$R_1^+(\pi/2,0)$\\
 &$R_1^+(\pi/\sqrt{2},0)$&\\
 \hline
  Ramsey II & $R_1(\pi/2,\pi)$&$R_1(\pi/2,(1/\sqrt{2}-1)\pi)$\\
\hline
SWAP II & $R^+_2(\pi,\pi)$ &$R^+_2(\pi,\pi)$\\
\end{tabular}
 \caption{\label{tab:czpulses}Two sequences of laser pulses for implementing a CNOT gate
 operation. Laser pulses applied to the i-th ion on the carrier transition are denoted by
 $R_i(\theta,\phi)$ and pulses on the blue sideband transition by $R_i^+(\theta,\phi)$, where
 $\theta=\Omega\cdot t$ denotes the  pulse area in terms of the Rabi frequency $\Omega$, the pulse length t  and  its phase $\phi$ \cite{roos2004}.}
\end{table}

Due to systematic imperfections and decoherence the actual evolution in our
experiment will deviate from the ideal unitary evolution given in
(\ref{eq:cnotunitaryA}) and (\ref{eq:cnotunitaryB}). An important systematic
error in our setup is imperfect addressing. Every laser pulse which addresses
one of the ions also slightly affects the neighboring ion qubits, due to the
finite size of the laser beam focus. In terms of the ratio of Rabi frequencies
between the addressed and the neighboring ions, this error is on the order of
2-3 \%. Furthermore, due to decoherence the output state after application of
an operation will in general be a mixed state. The major sources of decoherence
in our experimental setup are fluctuations of the laser frequency and the
ambient magnetic field \cite{schmidtkalerjphysb2003}.

The experimentally realized quantum gate including error sources is properly
described by a completely positive map $\mathcal{E}$. For an input state
$\rho$, the output state $\mathcal{E}(\rho)$ can be written in the operator sum
representation \cite{Chuang1997} as
\begin{equation}
\mathcal{E}(\rho)=\sum_{m,n=0}^{4^N-1}\chi_{mn}\hat{A}_m\rho\hat{A}_n^{\dagger},
\label{eq:operatorsum}
\end{equation}
where N is the number of qubits and the $\hat{A}_m$ are operators forming a
basis in the space of $2^N\times 2^N$-matrices. The \textit{process matrix}
$\chi$ contains complete information about the investigated process including
the influence of the environment on the qubits. The matrix $\chi$ can be
experimentally obtained by employing quantum process tomography. This procedure
requires $4^N$ input states $\rho_{in,i}$ which are linearly independent. For
every input state, the output state
 $\mathcal{E}(\rho_{in,i})$ has to be determined by quantum
state tomography. From this set of data the process matrix $\chi$ can be
obtained by inverting the relation in (\ref{eq:operatorsum}). However, to avoid
unphysical results caused by quantum noise in the measurement process, we
employ an iterative maximum likelihood algorithm \cite{jezek2003} in order to
find the physical process $\mathcal{E}$ which most likely generated the
measured data set.
\begin{figure*}[ht]
\includegraphics[width=\linewidth]{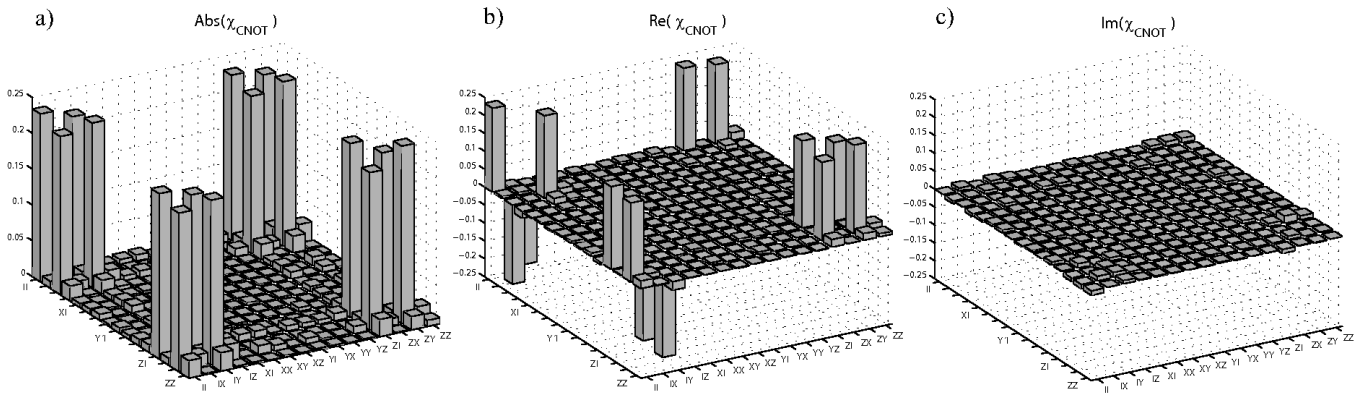}
\caption{Process matrix obtained by process tomography of a gate operation
implemented by pulse sequence (B) in Tab.\ \ref{tab:czpulses}. The absolute
value, real part and imaginary part of $\chi_{\text{CNOT}}$ are shown in a), b)
and c) respectively. The matrix is expressed in terms of the products of the
identity $\hat{I}$ and the Pauli operators $\hat{X}$, $\hat{Y}$ and $\hat{Z}$.
In order to compensate for the rotation $\hat{U}_Z$ the phase of all tomography
pulses was shifted appropriately.} \label{fig:gatetomo}
\end{figure*}
We choose the products of the single qubit states $|\psi_1\rangle=|S\rangle$,
$|\psi_2\rangle=|D\rangle$, $|\psi_3\rangle=(|D\rangle+i|S\rangle)/\sqrt{2}$
and $|\psi_4\rangle=(|D\rangle+|S\rangle)/\sqrt{2}$ as the 16  input states
necessary for a tomography of our two qubit quantum gates. Quantum state
tomography of a two qubit system then requires measurements in nine different
product state bases \cite{roos2004}. This results in a total of $16\times
9=144$ different measurement settings. Every expectation value is determined
through 100-250 individual experiments at a rate of 50 experiments/s which
requires 5-12 minutes of measurement time to gather all data necessary for the
estimation of $\chi$.

Quantum process tomography was carried out for the operations implemented by
pulse sequences (A) and (B) shown in Tab.\ \ref{tab:czpulses}. The duration of
the complete CNOT pulse sequence was $T_{\text{Gate}}=615\; \mu s$ for sequence
(A) and $T_{\text{Gate}}=502\; \mu s$ for (B), which is mainly determined by
the Rabi frequency on the blue sideband at $\Omega_{\text{BSB}}=2\pi\cdot4.4$
kHz. The resulting estimated process matrix $\chi_{\text{cnot}}$ for pulse
sequence (B) is shown in Fig.\ \ref{fig:gatetomo}.

\begingroup
\squeezetable
\begin{table}[hb]
\begin{tabular}{|c|c|c|c|c|c|c|c|}
  \hline
  & Seq. &$F_p$ in \% & $F_{\text{mean}}$ in \% & $\bar{S}_{\text{lin}}$ & max$(\Delta C)$ & Description \\
  \hline
  1&A & 88.8(7) & 91.0(6) & 0.20(1) & 0.86(2) & single gate $\mathcal{E}_{\text{CNOT(A)}}$ \\
  2&B & 90.8(6) & 92.6(6) & 0.17(1) & 0.84(2) & single gate $\mathcal{E}_{\text{CNOT(B)}}$\\
  \hline
  3&A & 87.7(7) & 90.1(6) & 0.21(1) & 0.81(2) & with pulse shaping \\
  4&A & 75(1) & 80(1) & 0.39(2) & 0.70(3) & no pulse shaping \\
  \hline
  5&AA & 79(1) & 83.4(8) & 0.34(2) & - &  $\mathcal{E}_{2\times\text{CNOT(A)}}$ \\
  6&A & 82.8 & 86.2 & 0.30 & - & $\mathcal{E}_{\text{CNOT(A)}}\circ \mathcal{E}_{\text{CNOT(A)}}$ \\
  7&BB & 72(1) & 77.4(8) & 0.41(1) & - & $\mathcal{E}_{2\times\text{CNOT(B)}}$ \\
  8&B & 79.8 & 83.8 & 0.34 & - & $\mathcal{E}_{\text{CNOT(B)}}\circ \mathcal{E}_{\text{CNOT(B)}}$ \\
  \hline
  \end{tabular}
  \caption{Performance of gate operations for different experimental
  realizations. The mean fidelity $F_{\text{mean}}$, the mean linear entropy
$\bar{S}_{\text{lin}}$ and the maximum change in entanglement max$(\Delta C)$
were inferred from an ensemble of $5\cdot 10^4$ random states. The quoted
errors are due to quantum projection noise \cite{roos2004}. For the results in
rows 1,2,5-8 the blue sideband Rabi frequency was set to
$\Omega_{\text{BSB}}=2\pi\cdot 4.4$ kHz. In rows 3,4 a higher Rabi frequency of
$\Omega_{\text{BSB}}=2\pi\cdot 5.3$ kHz was chosen. Rows 5,7 contain the
results for the tomographies of two concatenated gates. Additionally, rows 6,7
show the results predicted from the single gate analysis.} \label{tab:results}
\end{table}
\endgroup

The process matrix allows us to calculate various measures which characterize
the performance of the gate operation. We can directly calculate the
\textit{process fidelity}
$F_{p}=tr\left(\chi_{\text{id}}\cdot\chi_{\text{CNOT}}\right)$, which is the
overlap of our experimentally obtained $\chi_{\text{CNOT}}$ with the ideal
process matrix $\chi_{\text{id}}$ derived from the unitary evolution of the
gate given in (\ref{eq:cnotunitaryA}) and (\ref{eq:cnotunitaryB}),
respectively. Furthermore, using  Eq.\ (\ref{eq:operatorsum}) we can predict
the output state of the experimental gate operation for an arbitrary input
state. This enables us to investigate the gate performance for a large number
of numerically generated input states, similar to the analysis done for an
optical CNOT gate in Ref.\ \cite{obrien2004}. We do this by analyzing the
calculated output states in terms of their overlap
$F=\langle\psi_{\text{id,out}}|\mathcal{E}(\rho)|\psi_{\text{id,out}}\rangle$
with the ideal output states $|\psi_{\text{id,out}}\rangle$, their normalized
linear entropy $S_{\text{lin}}=4/3\cdot tr\left(1-\mathcal{E}(\rho)^2\right)$
and the change in entanglement from the input to the output states given by the
change in the concurrence, i.e.\ the difference $\Delta
C=C(\mathcal{E}(\rho))-C(\rho)$ \cite{Wootters1998PRL}. For $5\cdot 10^4$
randomly chosen pure input states drawn from the Haar measure on the unitary
group $U(4)$, we carry out such an analysis using the results of the process
tomography for the two types of quantum gates described above. We characterize
the gate performance by calculating the mean fidelity $F_{\text{mean}}$, the
mean linear entropy $\hat{S}_{\text{lin}}$ and by searching for the maximum
increase in entanglement max$(\Delta C)$. The results for a single gate
operation are shown in Tab.\ \ref{tab:results} (rows 1,2).
 Pulse sequence (B) shows a slightly better performance than sequence (A)
probably due to the 20 \% shorter phase gate sequence, which reduces the
influence of decoherence. Furthermore, errors due to imperfect addressing will
partially cancel in phase gate sequence (B), since the two $\pi/2$-pulses are
applied with a relative phase of $\pi$.

We used this kind of analysis to assess the influence of amplitude pulse
shaping on the gate performance. Exciting the blue sideband transition with
square laser pulses causes off-resonant excitation of the carrier transition
degrading the performance of the quantum gate. If higher gate speeds were to be
achieved by using higher sideband Rabi frequencies \cite{steane2000}, this
effect would become increasingly harmful. However, we largely suppress
off-resonant excitations by adiabatically switching on and off the laser pulses
\cite{wineland1998}. We demonstrated this by first carrying out a process
tomography of gate pulse sequence (B) for $T_{\text{gate}}=$ 520 $\mu s$ using
shaped pulses, that were adiabatically switched on and off with a rise and fall
time of 5 $\mu s$ (the standard setting for all reported results). Then we
carried out another gate tomography using simple square pulses. As can be seen
from the results in Tab.\ \ref{tab:results} (rows 3,4), the use of shaped
pulses considerably improves the gate performance compared with the result
\cite{schmidtkalernature2003} with square pulses.

\begin{figure}[hb]
\includegraphics[width=0.9\linewidth]{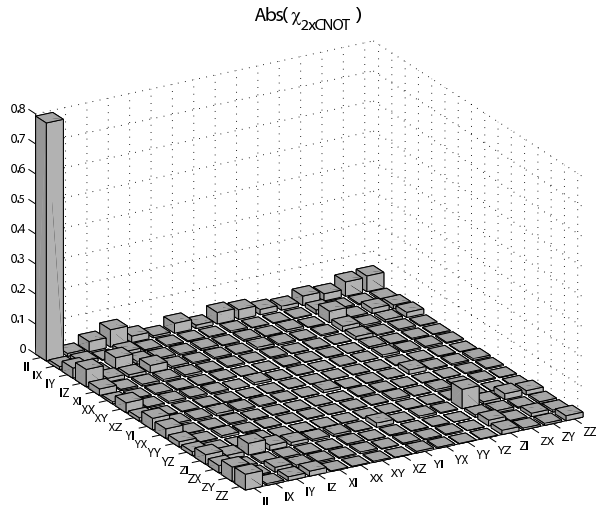}
\caption{Absolute value of the measured process matrix resulting from two gate
operations successively applied to a pair of ion qubits. Ideally the process
matrix should only contain the $II,II$-element. The height of the
$II,II$-element in the measured process matrix of 79.3 \% directly gives the
process fidelity.} \label{fig:2xgatetomo}

\end{figure}

Quantum algorithms will generally contain multiple quantum gates, which are
successively applied to a qubit register. The question arises whether the
performance of such a series of quantum gates can be inferred from the
knowledge of the single-gate performances in experimental implementations of an
algorithm. We investigated this issue by comparing the result of a process
tomography of two concatenated CNOT gates with the predictions inferred from
the single gate tomography, for the same set of experimental parameters.
Ideally, two concatenated CNOT yield the identity, $(U_{\text{CNOT}})^2=I$. The
measured process matrix $\chi_{\text{2xCNOT}}$ is shown in Fig.\
\ref{fig:2xgatetomo}. As expected, the dominant element of
$\chi_{\text{2xCNOT}}$ is the ${II,II}$-element. We determined the process
fidelity, mean fidelity and mean entropy of the operation described by
$\chi_{\text{2xCNOT}}$. For comparison, we calculated the same quantities for
the process $\mathcal{E}_{\text{CNOT}}\circ \mathcal{E}_{\text{CNOT}}$, where
$\mathcal{E}_{\text{CNOT}}$ was obtained from the single gate process
tomography. The results are shown in Tab. \ref{tab:results} (rows 5-8). In
general, one would expect both methods to yield the same results if the
dynamics of interaction between the qubits and the environment was Markovian,
thus producing uncorrelated errors in both gates. In our experiment, we
attribute the observed discrepancy to low-frequency magnetic noise giving rise
to magnetic fields that are constant over the course of the double gate
sequence but vary from experiment to experiment. In addition to characterizing
the performance of gates within larger blocks, concatenating quantum gates
might be useful for amplifying tiny errors in high-quality gates to a
measurable size.

The presented work demonstrates quantum process tomography to be a valuable
tool for assessing the performance of the fundamental operations of ion trap
quantum computers. Since a complete tomographic data set can be taken in a
comparably short amount of time, it is easily possible to compare the
performance of quantum gates for different pulse sequences and experimental
parameters. This has been illustrated by showing that adiabatically switching
on and off laser pulses instead of using rectangular shaped pulses make higher
gate speeds possible while preserving high gate fidelities. This technique and
careful optimization of the experimental parameters using process tomography
helped to significantly improve the gate fidelity compared with earlier results
\cite{schmidtkalernature2003,schmidtkalerapplphysb2003}. The results of the
concatenated gate tomography demonstrate the importance of analyzing
experimental quantum gate implementations not only as isolated objects but also
within a larger gate sequence. We expect that this technique will have
considerable impact on estimating the overall performance of future quantum
computers.

We acknowledge support by the Austrian Science Fund (FWF), by the European
Commission (QGATES, SCALA, CONQUEST networks) and by the Institut f\"ur
Quanteninformation GmbH. This material is based on work supported in part by
the US Army Research Office. K.~Kim acknowledges funding by the Lise-Meitner
program of the FWF. We acknowledge P.~Pham's contribution to the development of
a source of shaped RF pulses. H.~H\"affner acknowledges useful discussions with
H.~De~Raedt.


\end{document}